\documentclass[twocolumn,prb,aps,showpacs]{revtex4}

\usepackage{graphicx}
\usepackage{dcolumn}
\usepackage{amsmath}
\newlength{\figwidth}
\newlength{\figwidthb}
\begin{document}


\title{Reentrant spin glass transition in LuFe$_2$O$_{4-\delta}$}
\author{Fan Wang}
\affiliation{Department of Physics, University of Toronto, Toronto,
Ontario M5S~1A7, Canada}
\author{Jungho Kim}
\affiliation{Department of Physics, University of Toronto, Toronto,
Ontario M5S~1A7, Canada}
\author{G. D. Gu}
\affiliation{Department of Condensed Matter and Material Science,
Brookhaven National Laboratory, Upton, New York, 11973}
\author{Young-June Kim}
\email{yjkim@physics.utoronto.ca}
\affiliation{Department of
Physics, University of Toronto, Toronto, Ontario M5S~1A7, Canada}

\date{\today}
\begin{abstract}
We have carried out a comprehensive investigation of magnetic
properties of LuFe$_2$O$_4$, using AC susceptibility, DC
magnetization and specific heat. A magnetic phase transition around
$\sim$236 K was observed with DC magnetization and specific heat
measurements, which is identified as a paramagnetic to ferrimagnetic
transition based on the nonlinear susceptibility data. Upon further
cooling below this temperature, we also observed highly relaxational
magnetic behavior: the DC magnetization exhibits history and time
dependence, and the real and imaginary part of the AC susceptibility
shows large frequency dependence. Dynamic scaling of the AC
susceptibility data suggests that this low temperature phase can be
described as a reentrant spin glass phase. We also discuss magnetic
field dependence of the spin glass transition and aging, memory and
rejuvenation effect below the glass transition temperature around
228 K.

\end{abstract}


\maketitle

\section{Introduction}
Geometrical frustration plays an important role in determining
ground states and phase transitions in magnetic systems. A
triangular lattice in two-dimension in particular is one of the
simplest systems to study the effect of geometrical frustration.
LuFe$_2$O$_4$ is a member of RFe$_2$O$_4$ family of compounds, where
R can be Y, Ho, Er, Tm, Yb, and Lu. \cite{1} These materials all
have hexagonal layered structure, in which Fe ions form a triangular
lattice within each bilayer. \cite{2} Since the average charge
valence of Fe in this compound is +2.5, this system is expected to
exhibit charge order behavior similar to Fe$_3$O$_4$ \cite{30,31} or
half doped manganites. \cite{32} However, due to the geometrical
frustration introduced by the triangular lattice, understanding
charge order in this material is not straightforward. \cite{3}
Previous electron and x-ray diffraction studies have shown that
charge ordering sets in below $\sim$ 300K, and anomalous dielectric
dispersion was observed in this temperature range. \cite{3,4} In
particular, Ikeda and coworkers argued that the observed
pyroelectric signal below charge ordering temperature indicates
charge order driven ferroelectricity. \cite{4} This result has been
drawing much attention,\cite{56,57} since this would be the first
such observation of ferroelectricity with electronic origin. In
addition, it was observed that the pyroelectric signal shows an
unusual step around the spin ordering temperature, and a large
magnetodielectric response under low magnetic fields was also
observed in LuFe$_2$O$_4$ at room temperature,\cite{58} which
prompted further interests in this compound as a possible
multiferroic (or magnetic ferroelectric) material.

Although whether the magnetic and ferroelectric order parameters are
coupled in LuFe$_2$O$_4$ is not clear at the moment, LuFe$_2$O$_4$
exhibits quite interesting magnetic properties, as a result of the
geometrical frustration arising from the triangular lattice. Most of
earlier studies of the magnetism in RFe$_2$O$_4$ have been focused
on YFe$_2$O$_4$. Tanaka et al. first reported that Fe spins order
below 220 K based on their M\"{o}ssbauer experiments. \cite{5} In
their studies of transport properties, they also observed that there
are two distinct transitions at 240 K and 225 K, and the former
corresponds to Verwey-like charge ordering accompanied by magnetic
ordering. \cite{6} This was corroborated in the x-ray study of
Nakagawa and coworkers, in which first order structural phase
transitions were observed around these temperatures. \cite{7}
Recently, Ikeda et al. reported that more than two transitions exist
in YFe$_2$O$_4$ based on their x-ray powder diffraction studies.
\cite{8} They also argued that the transition at 250 K corresponds
to charge and spin ordering.

However, it was also realized that the oxygen non-stoichiometry in
YFe$_2$O$_4$ can cause significant changes in its magnetic
properties, while LuFe$_2$O$_4$ is believed to be free from such
oxygen non-stoichiometry problems. \cite{9} In their comprehensive
magnetization and neutron scattering work on LuFe$_2$O$_4$, Iida and
coworkers were able to elucidate unusual magnetic properties of this
compound. \cite{10} Specifically, they found that the system does
not show any long range three-dimensional magnetic order down to 4.2
K. Instead, they argued that the system at low temperatures consists
of ferrimagnetic clusters of various sizes, based on their
thermoremanent magnetization measurements. The ferrimagnetism in
this case arises due to the mixture of S=2 and S=5/2 spins. In
recent neutron scattering experiments, however, sharp magnetic Bragg
peaks were observed, suggesting existence of long-range magnetic
order.\cite{38,neutron-new} Therefore, the nature of the ground
state of LFO is still not understood well.

In this paper, we report our comprehensive study of magnetic
properties of LuFe$_2$O$_4$ using AC susceptibility, DC
magnetization and specific heat. We have observed two magnetic
transitions: The high temperature transition occurs at $\sim$236K,
and corresponds to the previously observed ferrimagnetic
transition.\cite{10,38} The signature of this transition is also
observed in our specific heat measurements. In addition to this
ferrimagnetic transition, we observe an unusual magnetic transition
at a lower temperature, which shows relaxational behavior similar to
that of a spin-glass phase.

This paper is organized as follows. In the next section, we will
explain our sample preparation and characterization in detail. In
Sec. III , our experimental results from magnetic susceptibility and
specific heat measurements are presented. In Sec. IV, we will
discuss the implication of the observed results, and possible
connection with the charge order and ferroelectricity.


\section{Experimental Details}
LuFe$_2$O$_4$ (LFO) single crystals were grown using the travelling
solvent floating zone method at Brookhaven National Laboratory
following the method reported in Ref \onlinecite{33}. Our
experiments were done using the crystals from the same batch without
any special annealing procedure. The chemical composition of one of
the crystals was examined with electron probe microanalysis
(EPMA)with beam size less than 1 micron. The Lu/Fe ratio was
analyzed at 25 randomly selected points on the sample surface. The
average Lu/Fe ratio was 1.98$\pm$0.02, and the mean deviation from
the average value was less than 1$\%$. This result shows that the Lu
and Fe is homogeneously distributed with almost stoichiometric
ratio. The oxygen contents of two other pieces were studied using
X-ray photoemission spectroscopy (XPS), revealing that the oxygen
content in one sample was higher than the other sample, suggesting
that there is a small but finite oxygen non-stoichiometry issue in
LuFe$_2$O$_4$. It turns out that the magnetic and structural
properties of LuFe$_2$O$_{4-\delta}$ depends very sensitively on the
oxygen stoichiometry. Detailed study of phase diagram is still in
progress, but we made sure that all the samples studied in this work
show the same magnetic properties. This ensures that the variation
of $\delta$ among the samples studied here is very small. The
largest piece with a shape of a rectangular parallelepiped ($3
\times 3 \times 1$ mm) was used in our magnetization studies. DC
magnetization and AC susceptibility measurements were done using
Quantum Design MPMS SQUID magnetometer. Specific heat measurements
on the same sample were carried out using thermal relaxation method
on Quantum Design PPMS.

\begin{figure}
\begin{center}
\includegraphics[angle=0,width=3in]{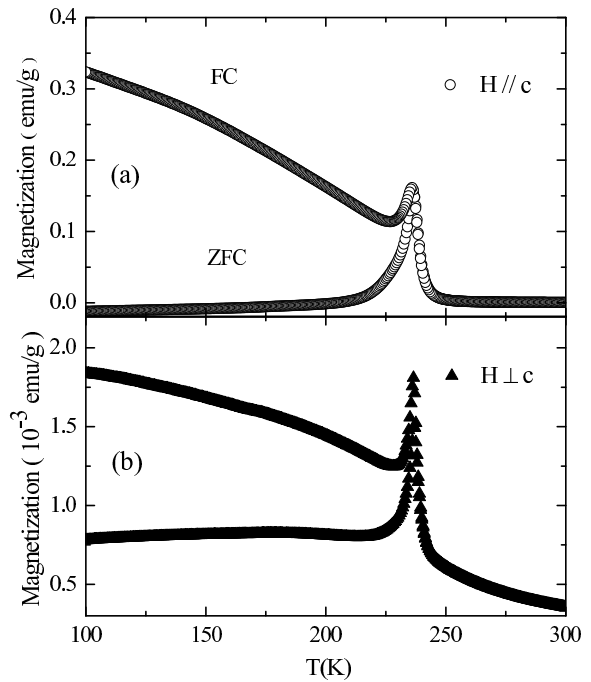}
\end{center}
\caption{Temperature dependence of magnetization measured with 10 Oe
field applied (a) parallel and (b) perpendicular to the c axis,
respectively. } \label{fig:Fig.1}
\end{figure}

\section{ Experimental Results}
\subsection{DC magnetization}
In Fig.~\ref{fig:Fig.1}(a), we show the temperature dependence of
the thermo-magnetization of LuFe$_2$O$_4$ obtained with 10 Oe field
applied parallel to the crystallographic c axis which is
perpendicular to the hexagonal planes. A sharp peak appears in the
magnetization curve at a temperature of $\sim$ 236K, below which the
field-cooled (FC) data begin to diverge from the zero-field-cooled
(ZFC) data. In Fig.~\ref{fig:Fig.1} (b), thermo-magnetization
obtained in a field perpendicular to the c axis is shown. Note that
the magnetization in this direction is 2 orders of magnitude smaller
than that shown in panel (a). This small magnetization can be
entirely accounted for by the possible sample misalignment with
respect to the field direction. This also illustrates that the easy
axis is along the c-axis, and the Ising anisotropy is very large.
The non-zero ZFC magnetization at low temperature in this case is
probably due to the small residual field in the magnetometer.

\begin{figure}
\begin{center}
\includegraphics[angle=-90,width=2.6in]{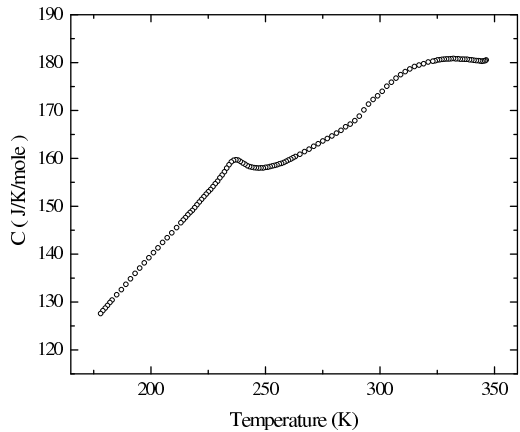}
\end{center}
\caption{ Temperature dependence of C (T) of sample A measured on
continuous cooling. }
 \label{fig:Fig.2}
\end{figure}

\subsection{Specific Heat}

Figure~\ref{fig:Fig.2} shows the temperature dependence of the
specific heat, $C(T)$, of the same sample used in the magnetization
study. One can identify two features in this curve. The high
temperature feature above 300K is relatively broad and has maximum
at $\sim$ 330K. This feature is probably related to the 3D charge
order observed in previous electron and x-ray diffraction studies.
\cite{3} The low temperature peak emerges below $\sim$250K and has a
cusp at $\sim$237K. The peak position of this low temperature
feature is very close to the peak in magnetic susceptibility,
suggesting that this feature is related to the magnetic phase
transition.

\subsection{AC susceptibility}

\begin{figure}
\begin{center}
\includegraphics[angle=0,width=3.1in]{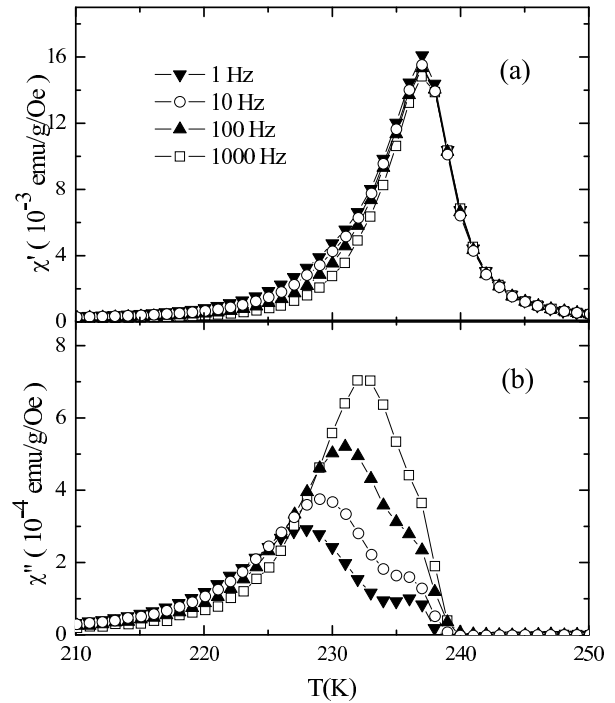}
\end{center}
\caption{Temperature dependence of the AC magnetic susceptibility of
sample A obtained with different frequencies as labeled. AC field
with amplitude $h_{ac}$=1 Oe was applied and the magnetization was
measured. The real and imaginary part of the susceptibility are
shown in part (a) and (b), respectively.}
 \label{fig:Fig.3}
\end{figure}

Figure~\ref{fig:Fig.3} shows the real and imaginary part of the AC
susceptibility as a function of temperature. The different curves
correspond to different driving frequencies of the AC field. The
amplitude of AC field was kept constant at $h_{ac}$=1 Oe. A
well-defined peak is observed for the real part of the
susceptibility $\chi'$ at 236 K and the low-temperature tail of this
peak decreases with increasing frequency. The imaginary part of the
susceptibility, $\chi''$, appears below $\sim$ 240 K, and consists
of two peaks. The high temperature component, appearing as a
shoulder, is located at $\sim$ 237 K and grows as frequency
increases, while the peak position remains the same. On the other
hand, the low temperature peak grows and shifts to higher
temperature with increasing frequency. Such a behavior is commonly
observed in spin glass systems.

\begin{figure}
\begin{center}
\includegraphics[angle=-90,width=3in]{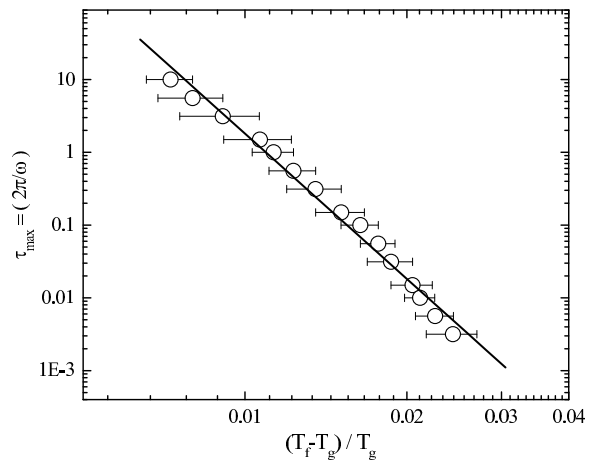}
\end{center}
\caption{The dynamic scaling of the reduced temperature vs
$\tau_{max}(T_f)=2\pi / \omega$ in a log-log scale for sample A. The
solid line is Eq.~\ref{equ:n} with $z\nu=6.6$, $\tau_0=1\times
10^{-13}$ s and $T_g=228.5$ K}
\label{fig:Fig.4}
\end{figure}

For a spin glass system, with decreasing temperature the spin
dynamics become sluggish, so that it takes longer time for a spin to
relax, and the maximum relaxation time increases accordingly. When
an external AC magnetic field with a driving frequency $\omega /
2\pi$ is applied to a spin glass system, if the maximum relaxation
time $\tau$$_{max}$ is longer than $\omega / 2\pi$, the system will
not be able to keep up with the oscillating field and become out of
equilibrium. Therefore, one can define the freezing temperature,
$T_f$, as the temperature, at which $\tau_{max}=2\pi /\omega$. As a
result, $T_f$ is a function of driving frequency $\omega$.
Experimentally $T_f (\omega$) can be determined either from the
maximum of $\chi'(\omega$) or from the inflection point of $\chi''
(\omega)$.\cite{23,29} Since the maximum of $\chi'$ is difficult to
identify due to the second peak located at 236K, we use the
inflection point of $\chi'' (\omega)$ to determine $T_f$. The
maximum relaxation time and $T_f (\omega)$ can be modeled with
conventional critical slowing down \cite{13}
\begin{eqnarray}
\tau_{max}=\tau_0(T_f/T_g-1)^{-z\nu},\label{equ:n}
\end{eqnarray}
where $T_g$ is the spin-glass transition temperature, $z$ is the
dynamical exponent, $\nu$ is the usual critical exponent for the
correlation length and $\tau$$_0$ is the microscopic flipping time
of the fluctuating spins. The scaling of the AC susceptibility is
shown in Fig.~\ref{fig:Fig.4}, and the best fit to Eq.~(\ref{equ:n})
yields $T_g=228.5 \pm 0.5$ K, $z\nu=6.6 \pm 1.1$ and $\tau_0
=10^{-13.0 \pm 1.6}$ s. The value of $\tau_0$ is very close to the
microscopic spin flip time $\sim 10^{-13}$ seconds in other spin
glass systems. \cite{28,29} The value of $z\nu$ is within the range
of well-known spin-glasses such as CuMn(4.6 at.\%) ($z\nu$=5.5)
\cite{14} and CdCr$_2$(In)S$_4$ ($z\nu=7$). \cite{15} This value of
$z\nu$ is also close to the value obtained from numerical
simulations in three-dimensional (3D) Ising spin glasses.
\cite{16,17,18} This scaling analysis indicates that the low
temperature phase is quite possibly a spin-glass phase. Thus, taken
together with the heat capacity and the susceptibility data, LFO
seems to undergo a continuous phase transition from a Curie
paramagnetic phase to ferrimagnetically ordered phase at $\sim$ 236K
and then to a reentrant spin-glass phase below 228K.


\subsection{Nonlinear susceptibility}
\begin{figure}
\begin{center}
\includegraphics[angle=-90,width=2.8in]{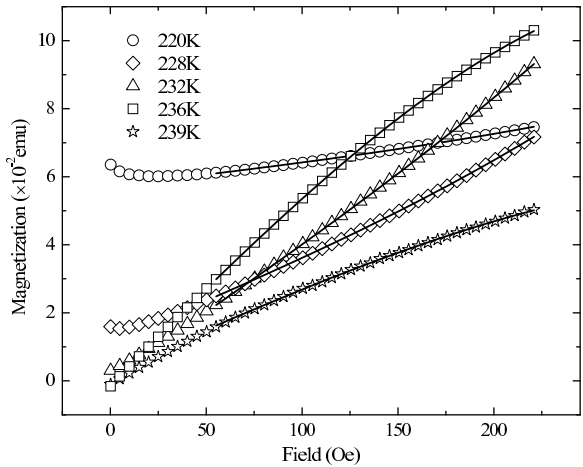}
\end{center}
\caption{M vs H curves of LFO at different fixed temperatures
measured after cooling under a 200 Oe magnetic field. The solid
lines are the fitting results using Eq.~\ref{equ:m}.}
\label{fig:Fig.5}
\end{figure}

Figure ~\ref{fig:Fig.5} shows the magnetization (M) versus field (H)
curves at different temperatures after cooling under a 200 Oe
magnetic field. At low temperatures, since the zero-field
magnetization is non-zero, there is thermal remanent magnetization
(TRM), which almost vanishes above $\sim$232K. At high temperatures,
there is no TRM and the slope of the M vs H curve is largely
determined by linear susceptibility, which shows maximum at 236 K
(see Fig.~\ref{fig:Fig.1}). At low temperatures, the magnetization
initially shows negative slope due to the relaxational behavior of
spin-glass phase.

Nonlinear susceptibility is a valuable tool to study the critical
behavior of a magnetic phase transition. In particular, the
spin-glass susceptibility is believed to be directly proportional to
the non-linear susceptibility.\cite{44,45,46} If a system undergoes
a SG phase transition, the magnetization M can be expanded by odd
order of H.

\begin{eqnarray}
M=M_0+a_{1}H+a_{3}H^{3}+....\label{equ:m}
\end{eqnarray}

\begin{figure}
\begin{center}
\includegraphics[angle=-90,width=3in]{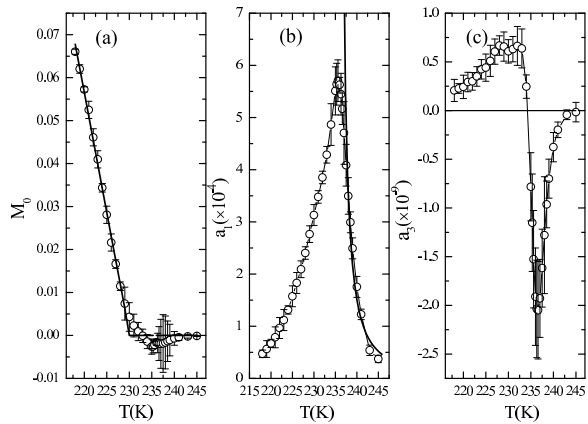}
\end{center}
\caption{(a) Constant offset $M_0$. The solid line is the fitting
using $(T_g -T)^{\beta}$. (b) Linear susceptibility $a_1$. The data
above $T_c=235.6$K is fitted by $(T-T_c)^{-\gamma}$, as shown by the
solid line. (c) The third order of nonlinear susceptibility $a_3$.
$M_0$, $a_1$ and $a_3$ were obtained by fitting the data in
Fig.~\ref{fig:Fig.5} into Eq.~\ref{equ:m}. } \label{fig:Fig.6}
\end{figure}

The data in Fig.~\ref{fig:Fig.5} (above 50 Oe) were fitted using
Eq.~(\ref{equ:m}), and the fitting results are shown as solid lines.
The constant offset $M_0$, linear susceptibility $a_1$ and the third
order nonlinear susceptibility $a_3$ obtained from the fits are
shown in Fig.~\ref{fig:Fig.6}. $M_0$ starts to grow below $\sim$235K
and then increases rapidly below $\sim$230K. The temperature at
which $M_0$ increases rapidly is close to $T_g$, suggesting that
$M_0$ behaves like the spin-glass order parameter. When the
temperature dependence of $M_0$ below 230K is fitted using $M_0 \sim
(T_g-T)^{\beta}$, best fits are obtained with $T_g=229$ K and
$\beta=1.0 \pm 0.2$, as shown by the solid line in
Fig.~\ref{fig:Fig.6}(a). This critical exponent $\beta$ is
consistent with that of canonical spin-glass systems.\cite{52,53}
The linear susceptibility peak in $a_1$ at $\sim$236K
(Fig.~\ref{fig:Fig.6}(b)) was also observed in DC magnetization and
heat capacity, and this indicates the existence of a magnetic phase
transition at $T_c \approx 236$K. If we fit the linear
susceptibility for $T>T_c$ to $a_1 \sim (T-T_c)^{-\gamma}$,
$\gamma=1.4 \pm 0.3$ is obtained, and the fitting result is plotted
as the solid line in Fig.~\ref{fig:Fig.6}(b). As temperature is
lowered, the third order nonlinear susceptibility $a_3$ shows a
negative minimum at 236K, and then it abruptly changes sign at 234K
and then show a broad positive peak at $\sim$230K. This behavior is
reminiscent of the recent observation that $a_3$ near a
ferromagnetic transition diverges negatively and positively as $T_c$
is approached from above and below, respectively.\cite{Nair03}
Therefore our observation of the zero-crossing of $a_3$ around 234K
seems to suggest that this high temperature magnetic transition is
described as ferrimagnetic (FM) transition. This assignment is also
consistent with earlier studies.\cite{10} The behavior of $a_3$
around 230K, however, is quite unusual for a spin-glass system. It
was observed that the non-linear susceptibility ($a_3$) is negative
with a cusp at the paramagnet to spin glass transition in Ref. 25
and 30. We observe broad positive peak around $T_g$, which may be
due to the reentrant nature of the spin-glass phase in LFO. In fact,
similarly unusual behavior have been observed in other reentrant
spin-glass systems.\cite{50,51} Our nonlinear susceptibility data
also supports the phase diagram suggested by AC susceptibility data.

\subsection{Non-equilibrium phenomena}

\begin{figure}
\begin{center}
\includegraphics[angle=-90,width=3in]{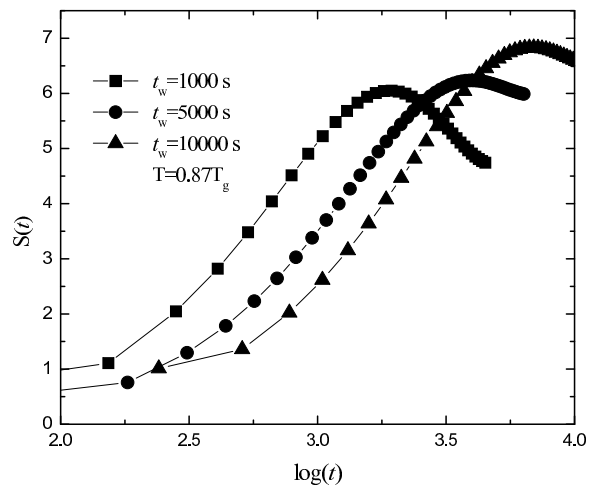}
\end{center}
\caption{Relaxation rate $S$ defined in the text is plotted as a
function of $\log(t)$ at $T=0.87T_g$ ($T_g=228.5$ K). Each curve is
obtained by measuring at $H=10$ Oe after waiting for $t_w$ following
the cool down.}
 \label{fig:Fig.7}
\end{figure}

Although spins are considered ``frozen'' in the spin glass phase,
due to the slow dynamics, the system simply does not  reach the
equilibrium state within the experimental time scale. As a result,
the spin glass system exhibits non-equilibrium phenomena. One such
example is aging. When a spin-glass system is cooled below $T_g$,
the spin-glass domain grows. Since this domain growth occurs
logarithmically in time, it is customary to define the relaxation
rate $S \equiv (1/H){\partial M/\partial
\log(t)}$.\cite{Fischer-book} In Fig.~\ref{fig:Fig.7}, we show our
data for $S(t)$ as a function of $\log(t)$. Note that the sample was
cooled down to 0.87$T_g$ $\sim$ 200 K in the absence of magnetic
field. After waiting for a certain time ($t_w$=1000s, 5000s and
10000s) without external field, the magnetization was recorded as a
function of time after a 10 Oe magnetic field was applied. As can
been seen from the figure, $t$ at which the maximum relaxation rate
occurs increases with increasing $t_w$, and in fact it is almost
equal to $t_w$. This kind of aging behavior illustrates
non-equilibrium dynamics of domain growth, and has been observed in
other spin glass systems.\cite{29,35}

\begin{figure}
\begin{center}
\includegraphics[angle=-90,width=3in]{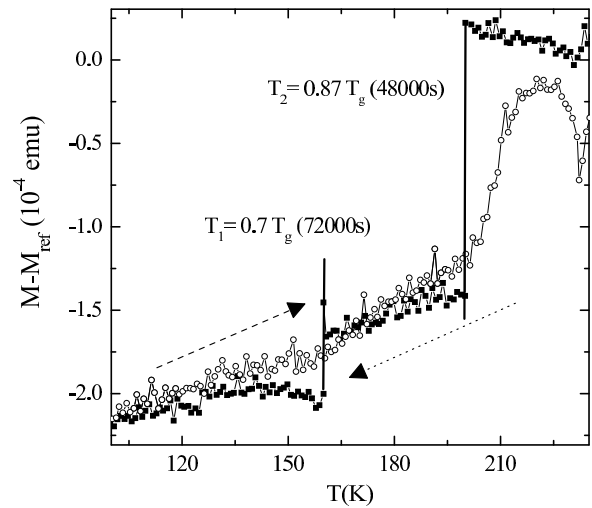}
\end{center}
\caption{ The relative magnetization $M-M_{ref}$ is plotted as a
function of temperature. The magnetization measured on continuous
cooling in $H=10$ Oe field, is plotted as solid symbols. During the
cooling, there were two halts at $T_1=0.7T_g$ and $T_2=0.87T_g$
($T_g=228$ K). The open symbols denote the measurements done on the
reheating. The reference $M_{ref}$ was obtained by continuous
cooling and reheating in 10 Oe. The cooling and heating rate in both
measurements were 2 K/min.}
 \label{fig:Fig.8}
\end{figure}

Another interesting example of non-equilibrium dynamics of
spin-glass system is the so-called memory effect. In order to show
this effect, we have measured temperature dependence of $M(T)$ in
two distinct routes. The first, $M_{ref}$, was obtained by cooling
in 10 Oe magnetic field from 300 K down to 50 K at a constant
cooling rate of 2 K/min and then heating back continuously at the
same rate. In the second route, M was recorded on cooling in 10 Oe
at the same rate from 300 K to 50 K with two halts at $T_1=160$ K
for 72000 s and at $T_2=200$ K for 48000 s. During the halts, the
external field is turned off to let the magnetization relax. After
each halt, M shows a clear deviation from the reference as
illustrated in Fig.~\ref{fig:Fig.8}, due to aging. After reaching 50
K, the sample temperature is increased continuously at 2 K/min rate
in $H = 10$ Oe. During the reheating, the system exhibits a
step-like feature at both $T_1$ and $T_2$. The jump at $T_1$ is not
very pronounced, but clear jump in $M(T)$ around $T_2$ is clearly
visible. This suggests that the system somehow remembers the history
of halts during cooling. Exceeding the halt points, $M$ recovers to
the reference value and the system is called rejuvenated. Such
aging, memory and rejuvenation behavior was observed in other spin
glass systems as well.\cite{28,42,43} These observations also
suggest that the low temperature phase of LFO is consistently
described as a spin-glass.

\subsection{Magnetic field dependence}

\begin{figure}
\begin{center}
\includegraphics[angle=0,width=2.9in]{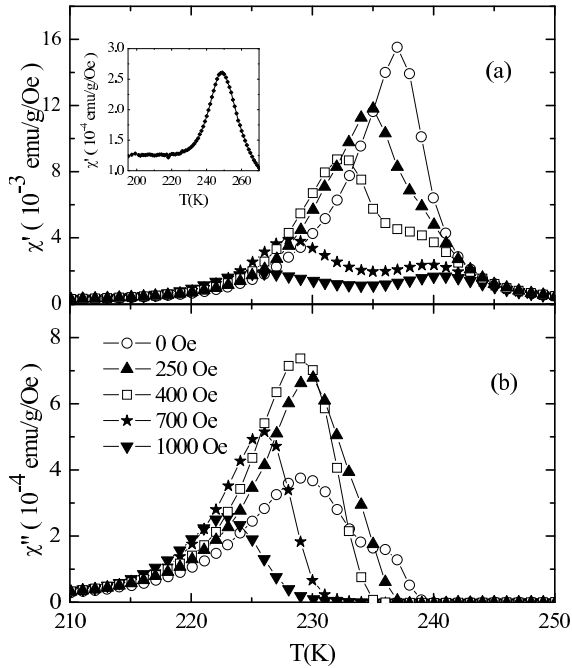}
\end{center}
\caption{ (a) The real and (b) the imaginary part of AC magnetic
susceptibility versus temperature. The driving frequency was fixed
at $\omega/2\ pi=10$ Hz and $h_{ac}$=1 Oe. Each curve was obtained
under different applied static  magnetic field of $H$. The inset
shows the data obtained with $H=1$ T.}
 \label{fig:Fig.9}
\end{figure}

In Fig.~\ref{fig:Fig.9}, AC susceptibility at 10 Hz driving
frequency is plotted as a function of temperature for different
external static magnetic fields H. As can be seen in the figure,
$\chi'$ is suppressed by the magnetic field. As the field increases,
the main peak of $\chi'$ decreases and a double-peak feature
emerges. The ferrimagnetic phase transition temperature determined
from specific heat and nonlinear susceptibility ($\sim$236 K) is
quite close to the position of the high temperature peak in $\chi'$,
which slightly increases with increasing field. The low temperature
peaks in $\chi'$ and $\chi''$ correspond to the spin-glass
transition. As the field increases, the low temperature peak in
$\chi'$ decreases and finally disappears under $\sim$ 1 T (as shown
in the inset). The peak in $\chi''$ also shifts to lower temperature
with increasing field. Under very high external field (above 1 T),
the spin-glass transition seems to be completely suppressed.

\begin{figure}
\begin{center}
\includegraphics[angle=-90,width=3in]{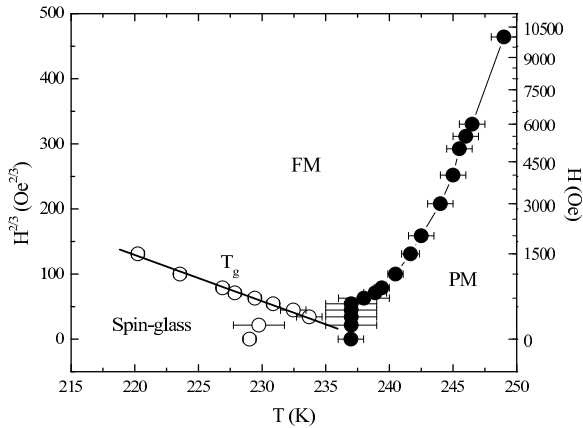}
\end{center}
\caption{Field vs. Temperature phase diagram. In order to show AT
line, we plot $H^{2/3}$ versus $T_g (\omega)$. The open and closed
symbols denote the low and high temperature transition temperatures
as described in the text. The thick solid line is the linear fit to
the AT line (Eq.~\ref{equ:p})} \label{fig:Fig.10}
\end{figure}

Figure~\ref{fig:Fig.10} shows the field versus temperature phase
diagram. Due to the large uncertainties associated with determining
transition temperature of two nearby phase transitions, the error
bars at low field is relatively large. However, the field dependence
of the PM-FM transition is very weak in this region, while it is
clear that the transition temperature gradually increases with
increasing field at high field. The spin-glass transition
temperature also exhibits substantial field dependence. The SG-FM
transition temperature is suppressed rapidly as a small field is
applied. In addition, a threshold field $h_0(\omega)\sim 200$ Oe is
observed, below which $T_g$ does not show a systematic change with
the change of the field. According to the mean field theory, there
exists a phase boundary in $H-T$ phase diagram called
de~Almeida-Thouless (AT) line,\cite{19} whereby a spin-glass phase
can only exist under this boundary (in the low field region). The AT
line is given by \cite{19}

\begin{eqnarray}
H\propto{(1-\frac{T_{g}(H)}{T_{g}(0)})^{3/2}}, \label{equ:p}
\end{eqnarray}

Here $H$ is the external magnetic field and $T_g(H)$ is the
field-dependent glass transition temperature.\cite{22} Our data fits
this relation very well as shown in Fig.~\ref{fig:Fig.10}, in which
a linear relationship between $T_g$ and $H^{2/3}$ is clearly
illustrated.


\section{Discussion}
We have presented a variety of experimental evidences showing that
the LFO sample goes through re-entrant spin glass transition around
228 K. However, this observation is quite puzzling in several
aspects. The first is the microscopic origin of the spin-glass
behavior. Conventionally, disordered spin arrangements or
interactions (random site or random bond) are necessary to produce
magnetic frustration required for spin-glass behavior. However, this
system, LFO, is considered highly stoichiometric, and in addition,
there exists charge ordering below 300 K, which implies that the
arrangement of Fe$^{2+}$ and Fe$^{3+}$ spins are regular. Therefore,
this system seems to possess only geometrical frustration as a
necessary ingredient for spin glass behavior. Unless spin-glass
behavior can arise from pure geometrical frustration, one must find
the missing ``disorder'' in this system to explain the observed
spin-glass behavior.

Before discussing the disorder effect in this system, it is useful
to examine magnetic interactions present. In LFO, both the Fe$^{2+}$
and Fe$^{3+}$ ions are in their high spin configuration, with the
spin angular momentum S=2 and S=5/2, respectively. The exchange
interactions between the Fe$^{2+}$-Fe$^{2+}$ and Fe$^{3+}$-Fe$^{3+}$
are presumably antiferromagnetic through the superexchange path via
the oxygen ions. However, the magnetic interaction between the
Fe$^{2+}$ and Fe$^{3+}$ ions requires further consideration. Note
that the Fe$^{2+}$ is in $d^{6}$ configuration, and the Hund's rule
dictates that the extra electron in this ion, compared to the
Fe$^{3+}$ ($d^{5}$) ion should point in the opposite direction of
the rest of the ``$d^{5}$'' electrons. Therefore, one can expect the
interaction between the Fe$^{2+}$ and Fe$^{3+}$ to be ferromagnetic
based on a kinetic energy argument, analogous to the double exchange
mechanism in manganites.\cite{54} However, since this compound is
insulating at all temperatures, such ``extra'' electron cannot be
mobile, but presumably resides in a resonance state between the two
neighboring Fe ions.

One of the important degrees of freedom in this system, often
overlooked, is the orbital degrees of freedom. Since the Fe ions all
have trigonal bipyramidal crystal field environment, the 3d orbitals
will split into three levels. The highest energy level will be
$d_{z^{2}}$, and there will be two doubly degenerate orbitals
($d_{xy}$ and $d_{x^{2}-y^{2}}$; $d_{xz}$ and $d_{yz}$) at lower
energies. Following the Hunds rule, one can easily see that
Fe$^{3+}$ ion has $d^{5}$ configuration and is isotropic. However,
Fe$^{2+}$ ion has $d^{6}$ configuration and the lowest energy
orbitals will have orbital degeneracy. In order to break this
orbital degeneracy, there could be a cooperative Jahn-Teller
distortion of Fe$^{2+}$ ions, once the charge order sets in. Taken
together, it is conceivable that two neighboring Fe$^{2+}$ and
Fe$^{3+}$ ions form a ``dimer'', sharing a minority spin electron.
In fact, such bond dimerization scenario has been considered in
their study of mixed valence B-site Fe ions in Fe$_3$O$_4$ by Seo
and coworkers.\cite{55} Such bond dimerization will break the
orbital degeneracy, and make the dimer spins form a highly
frustrated Kagome lattice, which may be responsible for the observed
spin glass behavior. Further structural investigation of this system
is required to address this conjecture.

Another important issue is the oxygen non-stoichiometry. If there
exists oxygen non-stoichiometry, this will necessarily affect the
ratio of Fe$^{2+}$ and Fe$^{3+}$ spins, leading to disorder in
charge order, and possibly disorder in the exchange interactions. In
fact, our preliminary studies suggest that the oxygen contents seem
to change magnetic properties quite dramatically, possibly
explaining different magnetic behavior of $\rm LuFe_2O_4$ compounds
reported in the literature.

Another possibility is the charge-ordering model itself. If the
charge ordering at 300 K is not of second-order kind, and has some
relaxational component, such as charge glass type ordering, then
this will naturally affect the spin ordering part. For example, one
can imagine charge-ordered domains exhibiting relaxor type behavior
and influence dielectric and magnetic behavior at lower
temperatures. Our preliminary x-ray scattering experiments also
suggest that the charge ordering in this system is quite complicated
and has nontrivial temperature dependence.

Another surprising aspect of the spin-glass transition is its large
temperature scale. Almost all known spin-glass systems occur at low
temperatures. This is due to the fact that the lower critical
dimension of the spin glass transition is believed to be between 2
and 3. Since a 3D SG system is very close to the lower critical
dimension, its ordering temperature is very close to absolute zero.
In that sense this observation of spin glass behavior at
temperatures above 200 K is not only unusual, but very surprising
for a quasi 2D system. Again, if the glassy nature of the system
arises from the charge sector, this may provide a natural
explanation.


\section{Summary}

The magnetic properties of $\rm LuFe_2O_4$ single crystals were
investigated with DC magnetization and AC susceptibility. Based on
the dynamic scaling of AC susceptibility and the behavior of
non-linear susceptibility, it is suggested that $\rm LuFe_2O_4$ goes
through first a ferrimagnetic ordering at 236 K, and then
subsequently goes through a reentrant spin-glass transition at $\sim
228.5$ K. Typical properties of spin glass system, such as aging,
memory, and rejuvenation have been also observed in this low
temperature phase. The field dependence of the spin glass transition
temperature is described well by the well-known de~Almeida-Thouless
theory. It was also observed that the ferrimagnetic transition
temperature shows quite sizable field dependence. In order to
understand the origin of the spin glass behavior in this compound,
possibilities based on the frustration, oxygen non-stoichiometry,
and glassy charge ordering have been discussed.


\begin{acknowledgements}

We would like to thank  David Ellis, S. M. Shapiro, G. Xu, J.
Brittain, A. Gershon and H. Zhang for invaluable discussions. The
work at University of Toronto was supported by Natural Sciences and
Engineering Research Council of Canada, Canadian Foundation for
Innovation, Ontario Innovation Trust, and Early Researcher Award by
Ontario Ministry of Research and Innovation. The work at Brookhaven
was supported by the U. S. Department of Energy, Office of Science.

\end{acknowledgements}


\begin{thebibliography}{44}
\expandafter\ifx\csname
natexlab\endcsname\relax\def\natexlab#1{#1}\fi
\expandafter\ifx\csname bibnamefont\endcsname\relax
  \def\bibnamefont#1{#1}\fi
\expandafter\ifx\csname bibfnamefont\endcsname\relax
  \def\bibfnamefont#1{#1}\fi
\expandafter\ifx\csname citenamefont\endcsname\relax
  \def\citenamefont#1{#1}\fi
\expandafter\ifx\csname url\endcsname\relax
  \def\url#1{\texttt{#1}}\fi
\expandafter\ifx\csname urlprefix\endcsname\relax\def\urlprefix{URL
}\fi \providecommand{\bibinfo}[2]{#2}
\providecommand{\eprint}[2][]{\url{#2}}

\bibitem[{\citenamefont{Kimizuka et~al.}(1990)\citenamefont{Kimizuka,
  Muromachi, and Siratori}}]{1}
\bibinfo{author}{\bibfnamefont{N.}~\bibnamefont{Kimizuka}},
  \bibinfo{author}{\bibfnamefont{E.}~\bibnamefont{Muromachi}},
  \bibnamefont{and} \bibinfo{author}{\bibfnamefont{K.}~\bibnamefont{Siratori}},
  \emph{\bibinfo{title}{Handbook on the Physics and Chemistry of Rare Earths}}
  (\bibinfo{publisher}{Elsevier Science}, \bibinfo{address}{Amsterdam},
  \bibinfo{year}{1990}), vol.~\bibinfo{volume}{13}, pp.
  \bibinfo{pages}{283--384}.

\bibitem[{\citenamefont{Kimizuka and Katsura}(1975)}]{2}
\bibinfo{author}{\bibfnamefont{N.}~\bibnamefont{Kimizuka}} \bibnamefont{and}
  \bibinfo{author}{\bibfnamefont{T.}~\bibnamefont{Katsura}},
  \bibinfo{journal}{J. Solid State Chem.} \textbf{\bibinfo{volume}{13}},
  \bibinfo{pages}{176} (\bibinfo{year}{1975}).

\bibitem[{\citenamefont{Hamilton}(1958)}]{30}
\bibinfo{author}{\bibfnamefont{W.~C.} \bibnamefont{Hamilton}},
  \bibinfo{journal}{Phys. Rev.} \textbf{\bibinfo{volume}{110}},
  \bibinfo{pages}{1050} (\bibinfo{year}{1958}).

\bibitem[{\citenamefont{Garc\'{\i}a and Sub\'{\i}as}(2004)}]{31}
\bibinfo{author}{\bibfnamefont{J.}~\bibnamefont{Garc\'{\i}a}} \bibnamefont{and}
  \bibinfo{author}{\bibfnamefont{G.}~\bibnamefont{Sub\'{\i}as}},
  \bibinfo{journal}{J. Phys: Condens. Matter} \textbf{\bibinfo{volume}{16}},
  \bibinfo{pages}{R145} (\bibinfo{year}{2004}).

\bibitem[{\citenamefont{Tomioka et~al.}(1995)\citenamefont{Tomioka, Asamitsu,
  Moritomo, Kuwahara, and Tokura}}]{32}
\bibinfo{author}{\bibfnamefont{Y.}~\bibnamefont{Tomioka}},
  \bibinfo{author}{\bibfnamefont{A.}~\bibnamefont{Asamitsu}},
  \bibinfo{author}{\bibfnamefont{Y.}~\bibnamefont{Moritomo}},
  \bibinfo{author}{\bibfnamefont{H.}~\bibnamefont{Kuwahara}}, \bibnamefont{and}
  \bibinfo{author}{\bibfnamefont{Y.}~\bibnamefont{Tokura}},
  \bibinfo{journal}{Phys. Rev. Lett.} \textbf{\bibinfo{volume}{74}},
  \bibinfo{pages}{5108} (\bibinfo{year}{1995}).

\bibitem[{\citenamefont{Yamada et~al.}(1997)\citenamefont{Yamada, Shinichiro,
  and Ikeda}}]{3}
\bibinfo{author}{\bibfnamefont{Y.}~\bibnamefont{Yamada}},
  \bibinfo{author}{\bibfnamefont{N.}~\bibnamefont{Shinichiro}},
  \bibnamefont{and} \bibinfo{author}{\bibfnamefont{N.}~\bibnamefont{Ikeda}},
  \bibinfo{journal}{J. Phys. Soc. Japan} \textbf{\bibinfo{volume}{66}},
  \bibinfo{pages}{3733} (\bibinfo{year}{1997}).

\bibitem[{\citenamefont{Ikeda et~al.}(2005)\citenamefont{Ikeda, Ohsumi, Ohwada,
  Ishii, Inami, Kakurai, Murakami, K.Yoshii, Mori, Horibe et~al.}}]{4}
\bibinfo{author}{\bibfnamefont{N.}~\bibnamefont{Ikeda}},
  \bibinfo{author}{\bibfnamefont{H.}~\bibnamefont{Ohsumi}},
  \bibinfo{author}{\bibfnamefont{K.}~\bibnamefont{Ohwada}},
  \bibinfo{author}{\bibfnamefont{K.}~\bibnamefont{Ishii}},
  \bibinfo{author}{\bibfnamefont{T.}~\bibnamefont{Inami}},
  \bibinfo{author}{\bibfnamefont{K.}~\bibnamefont{Kakurai}},
  \bibinfo{author}{\bibfnamefont{Y.}~\bibnamefont{Murakami}},
  \bibinfo{author}{\bibnamefont{K.Yoshii}},
  \bibinfo{author}{\bibfnamefont{S.}~\bibnamefont{Mori}},
  \bibinfo{author}{\bibfnamefont{Y.}~\bibnamefont{Horibe}},
  \bibnamefont{et~al.}, \bibinfo{journal}{Nature}
  \textbf{\bibinfo{volume}{436}}, \bibinfo{pages}{1136} (\bibinfo{year}{2005}).

\bibitem[{\citenamefont{Zhang et~al.}(2007)\citenamefont{Zhang, Yang, Ma, Tian,
  and Li}}]{56}
\bibinfo{author}{\bibfnamefont{Y.}~\bibnamefont{Zhang}},
  \bibinfo{author}{\bibfnamefont{H.~X.} \bibnamefont{Yang}},
  \bibinfo{author}{\bibfnamefont{C.}~\bibnamefont{Ma}},
  \bibinfo{author}{\bibfnamefont{H.~F.} \bibnamefont{Tian}}, \bibnamefont{and}
  \bibinfo{author}{\bibfnamefont{J.~Q.} \bibnamefont{Li}},
  \bibinfo{journal}{Phys. Rev. Lett.} \textbf{\bibinfo{volume}{98}},
  \bibinfo{pages}{247602} (\bibinfo{year}{2007}).

\bibitem[{\citenamefont{Xiang and Whangbo}(2007)}]{57}
\bibinfo{author}{\bibfnamefont{H.~J.} \bibnamefont{Xiang}} \bibnamefont{and}
  \bibinfo{author}{\bibfnamefont{M.-H.} \bibnamefont{Whangbo}},
  \bibinfo{journal}{Phys. Rev. Lett.} \textbf{\bibinfo{volume}{98}},
  \bibinfo{pages}{246403} (\bibinfo{year}{2007}).

\bibitem[{\citenamefont{Subramanian et~al.}(2006)\citenamefont{Subramanian, He,
  Chen, Rogado, Calvarese, and Sleight}}]{58}
\bibinfo{author}{\bibfnamefont{M.~A.} \bibnamefont{Subramanian}},
  \bibinfo{author}{\bibfnamefont{T.}~\bibnamefont{He}},
  \bibinfo{author}{\bibfnamefont{J.}~\bibnamefont{Chen}},
  \bibinfo{author}{\bibfnamefont{N.~S.} \bibnamefont{Rogado}},
  \bibinfo{author}{\bibfnamefont{T.~G.} \bibnamefont{Calvarese}},
  \bibnamefont{and} \bibinfo{author}{\bibfnamefont{A.~W.}
  \bibnamefont{Sleight}}, \bibinfo{journal}{Adv. Mater.}
  \textbf{\bibinfo{volume}{18}}, \bibinfo{pages}{1737} (\bibinfo{year}{2006}).

\bibitem[{\citenamefont{Tanaka et~al.}(1979)\citenamefont{Tanaka, Kato,
  Kimizuka, and Siratori}}]{5}
\bibinfo{author}{\bibfnamefont{M.}~\bibnamefont{Tanaka}},
  \bibinfo{author}{\bibfnamefont{M.}~\bibnamefont{Kato}},
  \bibinfo{author}{\bibfnamefont{N.}~\bibnamefont{Kimizuka}}, \bibnamefont{and}
  \bibinfo{author}{\bibfnamefont{K.}~\bibnamefont{Siratori}},
  \bibinfo{journal}{J. Phys. Soc. Japan} \textbf{\bibinfo{volume}{47}},
  \bibinfo{pages}{1737} (\bibinfo{year}{1979}).

\bibitem[{\citenamefont{Tanaka et~al.}(1982)\citenamefont{Tanaka, Akimitsu,
  Inada, Kimizuka, Shindo, and Siratori}}]{6}
\bibinfo{author}{\bibfnamefont{M.}~\bibnamefont{Tanaka}},
  \bibinfo{author}{\bibfnamefont{J.}~\bibnamefont{Akimitsu}},
  \bibinfo{author}{\bibfnamefont{Y.}~\bibnamefont{Inada}},
  \bibinfo{author}{\bibfnamefont{N.}~\bibnamefont{Kimizuka}},
  \bibinfo{author}{\bibfnamefont{I.}~\bibnamefont{Shindo}}, \bibnamefont{and}
  \bibinfo{author}{\bibfnamefont{N.}~\bibnamefont{Siratori}},
  \bibinfo{journal}{Solid State Commun.} \textbf{\bibinfo{volume}{44}},
  \bibinfo{pages}{687} (\bibinfo{year}{1982}).

\bibitem[{\citenamefont{Nakagawa et~al.}(1979)\citenamefont{Nakagawa, Inazumi,
  Kimizuka, and Siratori}}]{7}
\bibinfo{author}{\bibfnamefont{Y.}~\bibnamefont{Nakagawa}},
  \bibinfo{author}{\bibfnamefont{M.}~\bibnamefont{Inazumi}},
  \bibinfo{author}{\bibfnamefont{N.}~\bibnamefont{Kimizuka}}, \bibnamefont{and}
  \bibinfo{author}{\bibfnamefont{K.}~\bibnamefont{Siratori}},
  \bibinfo{journal}{J. Phys. Soc. Japan} \textbf{\bibinfo{volume}{47}},
  \bibinfo{pages}{1369} (\bibinfo{year}{1979}).

\bibitem[{\citenamefont{Ikeda et~al.}(2002)\citenamefont{Ikeda, Mori, Kohn,
  Mizumaki, and Akao}}]{8}
\bibinfo{author}{\bibfnamefont{N.}~\bibnamefont{Ikeda}},
  \bibinfo{author}{\bibfnamefont{R.}~\bibnamefont{Mori}},
  \bibinfo{author}{\bibfnamefont{K.}~\bibnamefont{Kohn}},
  \bibinfo{author}{\bibfnamefont{M.}~\bibnamefont{Mizumaki}}, \bibnamefont{and}
  \bibinfo{author}{\bibfnamefont{T.}~\bibnamefont{Akao}},
  \bibinfo{journal}{Ferroelectrics} \textbf{\bibinfo{volume}{272}},
  \bibinfo{pages}{309} (\bibinfo{year}{2002}).

\bibitem[{\citenamefont{Nakagawa et~al.}(1981)\citenamefont{Nakagawa, Kishi,
  Hiroyoshi, Kimizuka, and Siratori}}]{9}
\bibinfo{author}{\bibfnamefont{Y.}~\bibnamefont{Nakagawa}},
  \bibinfo{author}{\bibfnamefont{M.}~\bibnamefont{Kishi}},
  \bibinfo{author}{\bibfnamefont{H.}~\bibnamefont{Hiroyoshi}},
  \bibinfo{author}{\bibfnamefont{N.}~\bibnamefont{Kimizuka}}, \bibnamefont{and}
  \bibinfo{author}{\bibfnamefont{K.}~\bibnamefont{Siratori}},
  \emph{\bibinfo{title}{Ferrites, Pro. 3rd Int. Conf. Ferrites, Kyoto}}
  (\bibinfo{publisher}{CAPJ}, \bibinfo{address}{Tokyo}, \bibinfo{year}{1981}),
  p. \bibinfo{pages}{115}.

\bibitem[{\citenamefont{Iida et~al.}(1993)\citenamefont{Iida, Tanaka, Nakagawa,
  Funahashi, Kimizuka, and Takekawa}}]{10}
\bibinfo{author}{\bibfnamefont{J.}~\bibnamefont{Iida}},
  \bibinfo{author}{\bibfnamefont{M.}~\bibnamefont{Tanaka}},
  \bibinfo{author}{\bibfnamefont{Y.}~\bibnamefont{Nakagawa}},
  \bibinfo{author}{\bibfnamefont{S.}~\bibnamefont{Funahashi}},
  \bibinfo{author}{\bibfnamefont{N.}~\bibnamefont{Kimizuka}}, \bibnamefont{and}
  \bibinfo{author}{\bibfnamefont{S.}~\bibnamefont{Takekawa}},
  \bibinfo{journal}{J. Phys. Soc. Japan} \textbf{\bibinfo{volume}{62}},
  \bibinfo{pages}{1723} (\bibinfo{year}{1993}).

\bibitem[{\citenamefont{Nagai et~al.}()\citenamefont{Nagai, Matsuda, Ishii,
  Kakurai, Kito, Ikeda, and Yamada}}]{38}
\bibinfo{author}{\bibfnamefont{S.}~\bibnamefont{Nagai}},
  \bibinfo{author}{\bibfnamefont{M.}~\bibnamefont{Matsuda}},
  \bibinfo{author}{\bibfnamefont{Y.}~\bibnamefont{Ishii}},
  \bibinfo{author}{\bibfnamefont{K.}~\bibnamefont{Kakurai}},
  \bibinfo{author}{\bibfnamefont{H.}~\bibnamefont{Kito}},
  \bibinfo{author}{\bibfnamefont{N.}~\bibnamefont{Ikeda}}, \bibnamefont{and}
  \bibinfo{author}{\bibfnamefont{Y.}~\bibnamefont{Yamada}},
  \bibinfo{note}{unpublished}.

\bibitem[{\citenamefont{Christianson et~al.}()\citenamefont{Christianson,
  Lumsden, Angst, Yamani, Tian, Jin, Payzant, Nagler, Sales, and
  Mandrus}}]{neutron-new}
\bibinfo{author}{\bibfnamefont{A.}~\bibnamefont{Christianson}},
  \bibinfo{author}{\bibfnamefont{M.}~\bibnamefont{Lumsden}},
  \bibinfo{author}{\bibfnamefont{M.}~\bibnamefont{Angst}},
  \bibinfo{author}{\bibfnamefont{Z.}~\bibnamefont{Yamani}},
  \bibinfo{author}{\bibfnamefont{W.}~\bibnamefont{Tian}},
  \bibinfo{author}{\bibfnamefont{R.}~\bibnamefont{Jin}},
  \bibinfo{author}{\bibfnamefont{E.}~\bibnamefont{Payzant}},
  \bibinfo{author}{\bibfnamefont{S.}~\bibnamefont{Nagler}},
  \bibinfo{author}{\bibfnamefont{B.}~\bibnamefont{Sales}}, \bibnamefont{and}
  \bibinfo{author}{\bibfnamefont{D.}~\bibnamefont{Mandrus}},
  \eprint{arXiv:0711.3560v1}.

\bibitem[{\citenamefont{Iida et~al.}(1990)\citenamefont{Iida, Takekawa, and
  Kimizuka}}]{33}
\bibinfo{author}{\bibfnamefont{J.}~\bibnamefont{Iida}},
  \bibinfo{author}{\bibfnamefont{S.}~\bibnamefont{Takekawa}}, \bibnamefont{and}
  \bibinfo{author}{\bibfnamefont{N.}~\bibnamefont{Kimizuka}},
  \bibinfo{journal}{J. Cryst Growth} \textbf{\bibinfo{volume}{102}},
  \bibinfo{pages}{398} (\bibinfo{year}{1990}).

\bibitem[{\citenamefont{Mattsson et~al.}(1995)\citenamefont{Mattsson, Jonsson,
  Nordblad, ArugaKatori, and Ito}}]{23}
\bibinfo{author}{\bibfnamefont{J.}~\bibnamefont{Mattsson}},
  \bibinfo{author}{\bibfnamefont{T.}~\bibnamefont{Jonsson}},
  \bibinfo{author}{\bibfnamefont{P.}~\bibnamefont{Nordblad}},
  \bibinfo{author}{\bibfnamefont{H.} \bibnamefont{ArugaKatori}},
  \bibnamefont{and} \bibinfo{author}{\bibfnamefont{A.}~\bibnamefont{Ito}},
  \bibinfo{journal}{Phys. Rev. Lett.} \textbf{\bibinfo{volume}{74}},
  \bibinfo{pages}{4305} (\bibinfo{year}{1995}).

\bibitem[{\citenamefont{Gunnarsson et~al.}(1992)\citenamefont{Gunnarsson,
  Svedlindh, Andersson, Nordblad, Lundgren, ArugaKatori, and Ito}}]{29}
\bibinfo{author}{\bibfnamefont{K.}~\bibnamefont{Gunnarsson}},
  \bibinfo{author}{\bibfnamefont{P.}~\bibnamefont{Svedlindh}},
  \bibinfo{author}{\bibfnamefont{J.~O.} \bibnamefont{Andersson}},
  \bibinfo{author}{\bibfnamefont{P.}~\bibnamefont{Nordblad}},
  \bibinfo{author}{\bibfnamefont{L.}~\bibnamefont{Lundgren}},
  \bibinfo{author}{\bibfnamefont{H.} \bibnamefont{ArugaKatori}},
  \bibnamefont{and} \bibinfo{author}{\bibfnamefont{A.}~\bibnamefont{Ito}},
  \bibinfo{journal}{Phys. Rev. B} \textbf{\bibinfo{volume}{46}},
  \bibinfo{pages}{8227} (\bibinfo{year}{1992}).

\bibitem[{\citenamefont{Hohenberg and Halperin}(1977)}]{13}
\bibinfo{author}{\bibfnamefont{P.~C.} \bibnamefont{Hohenberg}}
  \bibnamefont{and} \bibinfo{author}{\bibfnamefont{B.~I.}
  \bibnamefont{Halperin}}, \bibinfo{journal}{Rev. Mod. Phys}
  \textbf{\bibinfo{volume}{49}}, \bibinfo{pages}{435} (\bibinfo{year}{1977}).

\bibitem[{\citenamefont{Mathieu et~al.}(2004)\citenamefont{Mathieu, Akahoshi,
  Asamitsu, Tomioka, and Y.Tokura}}]{28}
\bibinfo{author}{\bibfnamefont{R.}~\bibnamefont{Mathieu}},
  \bibinfo{author}{\bibfnamefont{D.}~\bibnamefont{Akahoshi}},
  \bibinfo{author}{\bibfnamefont{A.}~\bibnamefont{Asamitsu}},
  \bibinfo{author}{\bibfnamefont{Y.}~\bibnamefont{Tomioka}}, \bibnamefont{and}
  \bibinfo{author}{\bibnamefont{Y.Tokura}}, \bibinfo{journal}{Phys. Rev. Lett.}
  \textbf{\bibinfo{volume}{93}}, \bibinfo{pages}{227202}
  (\bibinfo{year}{2004}).

\bibitem[{\citenamefont{Mydosh}(1993)}]{14}
\bibinfo{author}{\bibfnamefont{J.~A.} \bibnamefont{Mydosh}},
  \emph{\bibinfo{title}{Spin Glass: An Experimental Introduction}}
  (\bibinfo{publisher}{Taylor $\&$ Francis}, \bibinfo{address}{London},
  \bibinfo{year}{1993}).

\bibitem[{\citenamefont{Vincent and Hammann}(1987)}]{15}
\bibinfo{author}{\bibfnamefont{E.}~\bibnamefont{Vincent}} \bibnamefont{and}
  \bibinfo{author}{\bibfnamefont{J.}~\bibnamefont{Hammann}},
  \bibinfo{journal}{J. Phys. C} \textbf{\bibinfo{volume}{20}},
  \bibinfo{pages}{2659} (\bibinfo{year}{1987}).

\bibitem[{\citenamefont{Campbell}(1988)}]{16}
\bibinfo{author}{\bibfnamefont{I.~A.} \bibnamefont{Campbell}},
  \bibinfo{journal}{Phys. Rev. B} \textbf{\bibinfo{volume}{37}},
  \bibinfo{pages}{9800} (\bibinfo{year}{1988}).

\bibitem[{\citenamefont{Ogielski}(1985)}]{17}
\bibinfo{author}{\bibfnamefont{A.~T.} \bibnamefont{Ogielski}},
  \bibinfo{journal}{Phys. Rev. B} \textbf{\bibinfo{volume}{32}},
  \bibinfo{pages}{7384} (\bibinfo{year}{1985}).

\bibitem[{\citenamefont{Ogielski and Morgenstern}(1985)}]{18}
\bibinfo{author}{\bibfnamefont{A.~T.} \bibnamefont{Ogielski}} \bibnamefont{and}
  \bibinfo{author}{\bibfnamefont{I.}~\bibnamefont{Morgenstern}},
  \bibinfo{journal}{Phys. Rev. Lett.} \textbf{\bibinfo{volume}{54}},
  \bibinfo{pages}{928} (\bibinfo{year}{1985}).

\bibitem[{\citenamefont{Ramirez et~al.}(1990)\citenamefont{Ramirez, Espinosa,
  and Cooper}}]{44}
\bibinfo{author}{\bibfnamefont{A.~P.} \bibnamefont{Ramirez}},
  \bibinfo{author}{\bibfnamefont{G.~P.} \bibnamefont{Espinosa}},
  \bibnamefont{and} \bibinfo{author}{\bibfnamefont{A.~S.}
  \bibnamefont{Cooper}}, \bibinfo{journal}{Phys. Rev. Lett.}
  \textbf{\bibinfo{volume}{64}}, \bibinfo{pages}{2070} (\bibinfo{year}{1990}).

\bibitem[{\citenamefont{Gingras et~al.}(1996)\citenamefont{Gingras, Stager,
  Gaulin, Raju, and Greedan}}]{45}
\bibinfo{author}{\bibfnamefont{M.~J.~P.} \bibnamefont{Gingras}},
  \bibinfo{author}{\bibfnamefont{C.~V.} \bibnamefont{Stager}},
  \bibinfo{author}{\bibfnamefont{B.~D.} \bibnamefont{Gaulin}},
  \bibinfo{author}{\bibfnamefont{N.~P.} \bibnamefont{Raju}}, \bibnamefont{and}
  \bibinfo{author}{\bibfnamefont{J.~E.} \bibnamefont{Greedan}},
  \bibinfo{journal}{J. Appl. Phys} \textbf{\bibinfo{volume}{79}},
  \bibinfo{pages}{6170} (\bibinfo{year}{1996}).

\bibitem[{\citenamefont{Rivadulla et~al.}(2004)\citenamefont{Rivadulla,
  L\'{o}pez-Quintela, and Rivas}}]{46}
\bibinfo{author}{\bibfnamefont{F.}~\bibnamefont{Rivadulla}},
  \bibinfo{author}{\bibfnamefont{M.~A.} \bibnamefont{L\'{o}pez-Quintela}},
  \bibnamefont{and} \bibinfo{author}{\bibfnamefont{J.}~\bibnamefont{Rivas}},
  \bibinfo{journal}{Phys. Rev. Lett.} \textbf{\bibinfo{volume}{93}},
  \bibinfo{pages}{167206} (\bibinfo{year}{2004}).

\bibitem[{\citenamefont{Bouchiat}(1986)}]{52}
\bibinfo{author}{\bibfnamefont{H.}~\bibnamefont{Bouchiat}},
  \bibinfo{journal}{J.Phys. (Paris)} \textbf{\bibinfo{volume}{47}},
  \bibinfo{pages}{71} (\bibinfo{year}{1986}).

\bibitem[{\citenamefont{Wakimoto et~al.}(2000)\citenamefont{Wakimoto, Ueki, Endoh, and Yamada}}]{53}
\bibinfo{author}{\bibfnamefont{S.}~\bibnamefont{Wakimoto}},
  \bibinfo{author}{\bibfnamefont{S.}~\bibnamefont{Ueki}},
  \bibinfo{author}{\bibfnamefont{Y.}~\bibnamefont{Endoh}}, \bibnamefont{and}
  \bibinfo{author}{\bibfnamefont{K.}~\bibnamefont{Yamada}},
  \bibinfo{journal}{Phys. Rev. B} \textbf{\bibinfo{volume}{62}},
  \bibinfo{pages}{3547} (\bibinfo{year}{2000}).

\bibitem[{\citenamefont{Nair and Banerjee}(2003)}]{Nair03}
\bibinfo{author}{\bibfnamefont{S.}~\bibnamefont{Nair}} \bibnamefont{and}
  \bibinfo{author}{\bibfnamefont{A.}~\bibnamefont{Banerjee}},
  \bibinfo{journal}{Phys. Rev. B} \textbf{\bibinfo{volume}{68}},
  \bibinfo{pages}{094408} (\bibinfo{year}{2003}).

\bibitem[{\citenamefont{Suzuki and Suzuki}(2006)}]{50}
\bibinfo{author}{\bibfnamefont{I.~S.} \bibnamefont{Suzuki}} \bibnamefont{and}
  \bibinfo{author}{\bibfnamefont{M.}~\bibnamefont{Suzuki}},
  \bibinfo{journal}{Phys. Rev. B} \textbf{\bibinfo{volume}{73}},
  \bibinfo{pages}{94448} (\bibinfo{year}{2006}).

\bibitem[{\citenamefont{Sato et~al.}(2001)\citenamefont{Sato, Ando, Ogawa,
  Morimoto, and Ito}}]{51}
\bibinfo{author}{\bibfnamefont{T.}~\bibnamefont{Sato}},
  \bibinfo{author}{\bibfnamefont{T.}~\bibnamefont{Ando}},
  \bibinfo{author}{\bibfnamefont{T.}~\bibnamefont{Ogawa}},
  \bibinfo{author}{\bibfnamefont{S.}~\bibnamefont{Morimoto}}, \bibnamefont{and}
  \bibinfo{author}{\bibfnamefont{A.}~\bibnamefont{Ito}},
  \bibinfo{journal}{Phys. Rev. B} \textbf{\bibinfo{volume}{64}},
  \bibinfo{pages}{184432} (\bibinfo{year}{2001}).

\bibitem[{\citenamefont{Fischer and Hertz}(1991)}]{Fischer-book}
\bibinfo{author}{\bibfnamefont{K.~H.} \bibnamefont{Fischer}} \bibnamefont{and}
  \bibinfo{author}{\bibfnamefont{J.~A.} \bibnamefont{Hertz}},
  \emph{\bibinfo{title}{Spin glasses}} (\bibinfo{publisher}{Cambridge
  University Press}, \bibinfo{address}{Cambridge, UK}, \bibinfo{year}{1991}).

\bibitem[{\citenamefont{Lundgren et~al.}(1985)\citenamefont{Lundgren, Nordblad,
  Svedlindh, and Beckman}}]{35}
\bibinfo{author}{\bibfnamefont{L.}~\bibnamefont{Lundgren}},
  \bibinfo{author}{\bibfnamefont{P.}~\bibnamefont{Nordblad}},
  \bibinfo{author}{\bibfnamefont{P.}~\bibnamefont{Svedlindh}},
  \bibnamefont{and} \bibinfo{author}{\bibfnamefont{O.}~\bibnamefont{Beckman}},
  \bibinfo{journal}{J. Appl. Phys.} \textbf{\bibinfo{volume}{57}},
  \bibinfo{pages}{3371} (\bibinfo{year}{1985}).

\bibitem[{\citenamefont{Vincent et~al.}(2000)\citenamefont{Vincent, Dupuis,
  Alba, Hammann, and Bouchaud}}]{42}
\bibinfo{author}{\bibfnamefont{E.}~\bibnamefont{Vincent}},
  \bibinfo{author}{\bibfnamefont{V.}~\bibnamefont{Dupuis}},
  \bibinfo{author}{\bibfnamefont{M.}~\bibnamefont{Alba}},
  \bibinfo{author}{\bibfnamefont{J.}~\bibnamefont{Hammann}}, \bibnamefont{and}
  \bibinfo{author}{\bibfnamefont{J.~P.} \bibnamefont{Bouchaud}},
  \bibinfo{journal}{Europhys. Lett.} \textbf{\bibinfo{volume}{50}},
  \bibinfo{pages}{674} (\bibinfo{year}{2000}).

\bibitem[{\citenamefont{Jonason et~al.}(1998)\citenamefont{Jonason, Vincent,
  Hammann, Bouchaud, and Nordblad}}]{43}
\bibinfo{author}{\bibfnamefont{K.}~\bibnamefont{Jonason}},
  \bibinfo{author}{\bibfnamefont{E.}~\bibnamefont{Vincent}},
  \bibinfo{author}{\bibfnamefont{J.}~\bibnamefont{Hammann}},
  \bibinfo{author}{\bibfnamefont{J.~P.} \bibnamefont{Bouchaud}},
  \bibnamefont{and} \bibinfo{author}{\bibfnamefont{P.}~\bibnamefont{Nordblad}},
  \bibinfo{journal}{Phys. Rev. Lett.} \textbf{\bibinfo{volume}{81}},
  \bibinfo{pages}{3243} (\bibinfo{year}{1998}).

\bibitem[{\citenamefont{de. Almeida and Thouless}(1978)}]{19}
\bibinfo{author}{\bibfnamefont{J.~R.~L.} \bibnamefont{de. Almeida}}
  \bibnamefont{and} \bibinfo{author}{\bibfnamefont{D.~J.}
  \bibnamefont{Thouless}}, \bibinfo{journal}{J. Phys. A}
  \textbf{\bibinfo{volume}{11}}, \bibinfo{pages}{983} (\bibinfo{year}{1978}).

\bibitem[{\citenamefont{Katori and Ito}(1994)}]{22}
\bibinfo{author}{\bibfnamefont{H.~A.} \bibnamefont{Katori}} \bibnamefont{and}
  \bibinfo{author}{\bibfnamefont{A.}~\bibnamefont{Ito}}, \bibinfo{journal}{J.
  Phys. Soc. Japan} \textbf{\bibinfo{volume}{63}}, \bibinfo{pages}{3122}
  (\bibinfo{year}{1994}).

\bibitem[{\citenamefont{Zener}(1951)}]{54}
\bibinfo{author}{\bibfnamefont{C.}~\bibnamefont{Zener}},
  \bibinfo{journal}{Phys. Rev.} \textbf{\bibinfo{volume}{82}},
  \bibinfo{pages}{403} (\bibinfo{year}{1951}).

\bibitem[{\citenamefont{Seo et~al.}(2002)\citenamefont{Seo, Ogata, and
  Fukuyama}}]{55}
\bibinfo{author}{\bibfnamefont{H.}~\bibnamefont{Seo}},
  \bibinfo{author}{\bibfnamefont{M.}~\bibnamefont{Ogata}}, \bibnamefont{and}
  \bibinfo{author}{\bibfnamefont{H.}~\bibnamefont{Fukuyama}},
  \bibinfo{journal}{Phys. Rev. B} \textbf{\bibinfo{volume}{65}},
  \bibinfo{pages}{85107} (\bibinfo{year}{2002}).

\end{thebibliography}

\end{document}